\documentclass[11pt,a4paper]{article}
\usepackage{eurosym}
\usepackage{amsfonts}
\usepackage{amsmath}
\usepackage{amsthm}
\usepackage{amssymb}
\usepackage{graphicx}
\usepackage{epstopdf}

\newif\iffigs\figstrue
\begin{document}

\begin{titlepage}
\vspace{.3cm} \vspace{1cm}
\begin{center}
{\Large \textsc{\ \\[0pt]
\vspace{5mm}Quantum Fluctuations and New Instantons I: \\[0pt]
Linear Unbounded Potential} }\\[0pt]

%\vspace{5mm}

\vspace{35pt} \textsc{V. F. Mukhanov$^{~a,b}$, E. Rabinovici$^{~c}$ and A.
S. Sorin$^{~d,e,f}$}\\[15pt]

{${}^{a}$ Ludwig Maxmillian University, \\[0pt]
Theresienstr. 37, 80333 Munich, Germany\\[0pt]
}e-mail: {\small \textit{mukhanov@physik.lmu.de}}\vspace{10pt}

{$^{b}${\small Korea Institute for Advanced Study\\[0pt]
Seoul, 02455, Korea}}\vspace{10pt}

{${}^{c}${\small Racah Institute of Physics, \\[0pt]
The Hebrew University of Jerusalem, 91904, Israel\\[0pt]
}}e-mail: {\small \textit{eliezer@vms.huji.ac.il}}\vspace{10pt}

{${}^{d}${\small Bogoliubov Laboratory of Theoretical Physics\\[0pt]
Joint Institute for Nuclear Research \\[0pt]
141980 Dubna, Moscow Region, Russia \\[0pt]
}}e-mail: {\small \textit{sorin@theor.jinr.ru}}\vspace{10pt}

{${}^{e}${\small National Research Nuclear University MEPhI\\[0pt]
(Moscow Engineering Physics Institute),\\[0pt]
Kashirskoe Shosse 31, 115409 Moscow, Russia}}\vspace{10pt}

{${}^{f}${\small Dubna State University, \\[0pt]
141980 Dubna (Moscow region), Russia}}\vspace{10pt}

\vspace{3mm}
\end{center}

%\vspace{1cm}

\begin{center}
\bf{Abstract}
\end{center}
We consider the decay of a false vacuum in circumstances where the methods suggested by Coleman run into difficulties.
We find that in these cases quantum fluctuations play a crucial role. Namely, they naturally induce both an ultraviolet and infrared cutoff scales, determined by the parameters of the classical solution, beyond which this solution cannot be trusted anymore. This leads to the appearance of a broad class of new $O(4)$ invariant instantons, which would have been singular in the absence of an ultraviolet cutoff. We apply our results to a case where the potential is unbounded from below in a linear way and in particular show how the problem of small instantons is resolved by taking into account the inevitable quantum fluctuations.

\end{titlepage}

\section{Introduction}

The study of the details of the fate of a false vacuum plays a key role in
understanding the properties of a variety of systems. It extends from the
understanding the characteristics of various phase transitions in condensed
matter and particle physics all the way to quantifying the human angst that
our universe itself may actually be in a state of a false vacuum. The gross
features of the decay mechanism, the formation of bubbles and the quantum
under the barrier tunneling involve both the classical and quantum aspects
of the problem. The description of tunneling in the non-relativistic quantum mechanics
for a particle moving in many dimensions was considered in \cite{Banks}.  In quantum field
theory the problem of a false vacuum instability was first addressed in \cite%
{Kobzarev}, assuming that the thickness of the wall is small compared to the
size of the bubble. Coleman \cite{Coleman} has generalized the results of 
\cite{Kobzarev} and suggested how to encapsulate and calculate the
non-perturbative quantum part of the process by identifying Euclidean
classical configurations which give the leading contribution to the
probability of the false vacuum decay. In this paper we study cases where
the method suggested by Coleman, as is, needs to be reexamined and modified.
In particular, we investigate how the theory must be modified by taking into
account the inevitable quantum fluctuations of the scalar field.

We start by a short review of the Coleman theory for the scalar field
potential $V\left( \varphi \right) ,$ which has a shape shown in Fig. \ref%
{Figure1} (for details see, for example, \cite{Weinberg}, \cite{Mukhanov1}%
)). The false vacuum configuration $\ \varphi \left( \mathbf{x}\right)
=\varphi _{0}$ is unstable and decays via bubble nucleation. The field
inside the emerged expanding bubble either tends to its value in the true
minimum or $\varphi $ $\rightarrow $ $-\infty $ for the unbounded potential.

\vspace{1.0cm}

\begin{figure}[hbt]
\begin{center}
%\iffigs
\includegraphics[height=60mm]{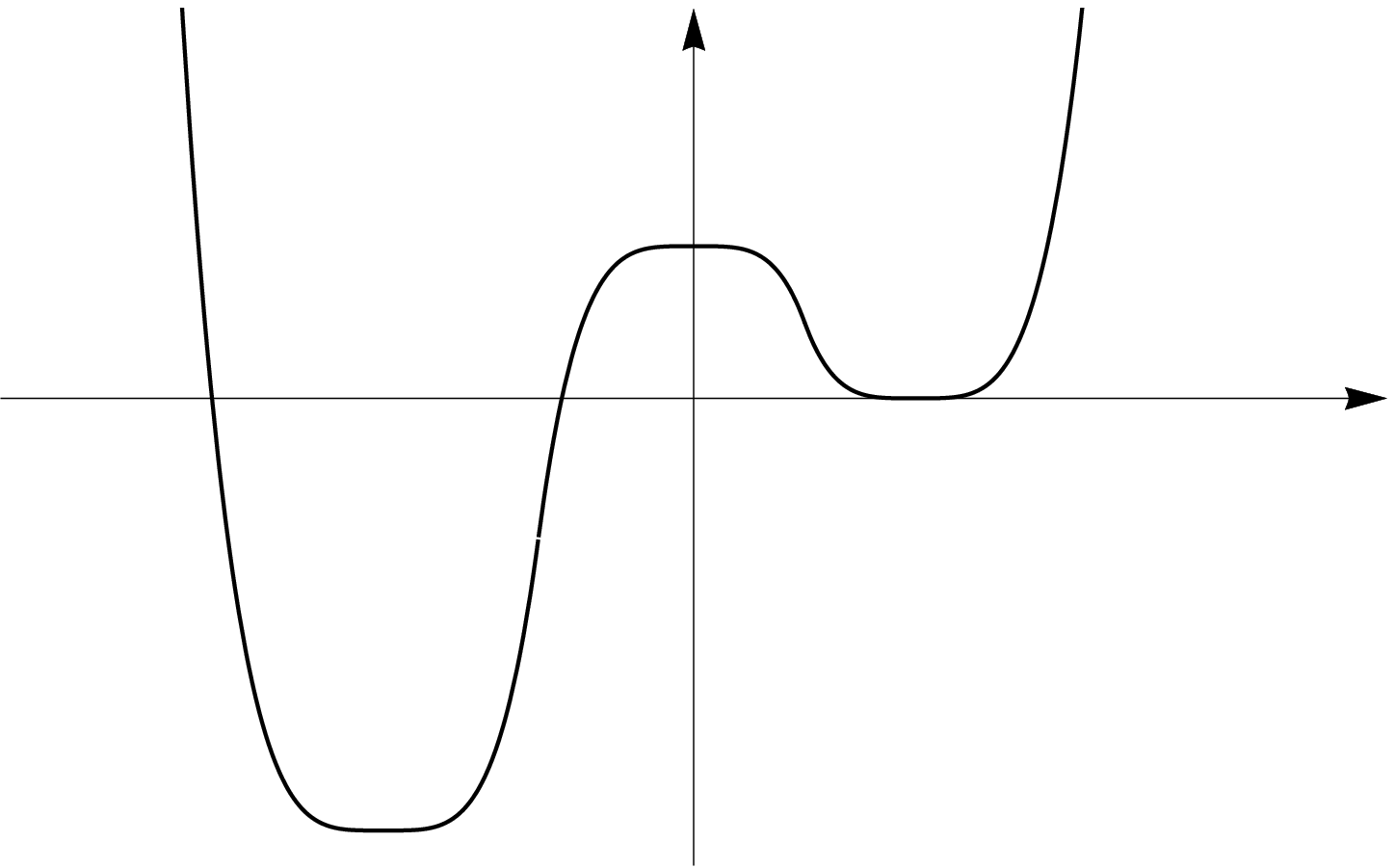} %\else
\end{center}
\par
\vspace*{-1.11cm} \hspace*{7.8cm} - $V_{min}$%
\par
\vspace*{-3.2cm} \hspace*{5.3cm} $\varphi_{min}$%
\par
\vspace*{-3.8cm} \hspace*{7.85cm}$V$%
\par
\vspace*{2.8cm} \hspace*{9.35cm}$\varphi_{0}$%
\par
\vspace*{-2.0cm} \hspace*{8.05cm}$V_{bar}$%
\par
\vspace*{0.8cm} \hspace*{13.0cm}$\varphi$%
\par
%%%%%%%%%%
%\fi
\vspace*{3cm} %\iffigs
%\hskip 1cm \unitlength=1.1mm
%\fi
\caption{}
\label{Figure1}
\end{figure}

\vspace{0.5cm}

The scalar field $\varphi \left( \mathbf{x},t\right) $ is a system with
infinitely many degrees of freedom with spatial coordinates $\mathbf{x=}%
\left( x^{1},x^{2},x^{3}\right) $ enumerating them, that is, $\varphi \left( 
\mathbf{x},t\right) =\varphi _{\mathbf{x}}\left( t\right) $. The scalar
field action can be written as 
\begin{equation}
S=\int (\mathcal{K-V)}\,dt\,,  \label{1}
\end{equation}%
where 
\begin{equation}
\mathcal{K=}\frac{1}{2}\int \left( \frac{\partial \varphi }{\partial t}%
\right) ^{2}\,d^{3}x\,,  \label{2}
\end{equation}%
and 
\begin{equation}
\mathcal{V=}\int \left( \frac{1}{2}\left( \frac{\partial \varphi }{\partial
x^{i}}\right) ^{2}+V\left( \varphi \right) \right) \,d^{3}x  \label{3}
\end{equation}%
are, correspondingly, the kinetic and interaction energies of the field
configuration $\varphi \left( \mathbf{x}\right) $ and its first time
derivatives$.$ Note that the functional $\mathcal{V}\left( \varphi \left( 
\mathbf{x}\right) \right) $ (\textit{but not} $V\left( \varphi \right) )$
plays the role of the potential energy when we are considering the
sub-barrier tunneling in quantum field theory. For the scalar field
potential $V\left( \varphi \right) ,$ shown in Fig. \ref{Figure1}, the field
configuration $\varphi \left( \mathbf{x}\right) =\varphi _{0}$ has zero
potential energy, $\mathcal{V}\left( \varphi _{0}\right) =0,$ while the
other homogeneous static solution $\varphi \left( \mathbf{x}\right) =\varphi
_{\min }$ has a lower energy $\mathcal{V}\left( \varphi _{\min }\right)
=V_{\min }\times $\textit{volume }$<$ $0.$ Therefore the local minimum at $%
\varphi _{0}$ is unstable (false vacuum) and decays via sub-barrier
tunneling as a result of which the bubbles of the true vacuum are formed. To
calculate the decay rate Coleman has assumed that in the semiclassical
approximation the dominant contribution to the tunneling rate comes from the
instanton - Euclidean solution of the scalar field equation, which matches
the metastable vacuum $\varphi \left( \mathbf{x}\right) =\varphi _{0}$ to
some classically allowed configuration $\varphi \left( \mathbf{x}\right) $
with $\mathcal{V}\left( \varphi \left( \mathbf{x}\right) \right) =0,$
describing the emerging bubble in Minkowski space. On symmetry grounds it
was shown \cite{Coleman1} that the minimal action has a $O\left( 4\right) $%
-invariant solution of the scalar field equation 
\begin{equation}
\frac{\partial ^{2}\varphi }{\partial \tau ^{2}}+\Delta \varphi -\frac{dV}{%
d\varphi }=0\,,  \label{4}
\end{equation}%
for which $\varphi $ depends only on $\varrho =\sqrt{\tau ^{2}+\mathbf{x}^{2}%
}$ and where $\tau =it$ is the Euclidean time. This reduces to the ordinary
differential equation 
\begin{equation}
\ddot{\varphi}(\varrho )+\frac{3}{\varrho }\,\dot{\varphi}(\varrho )-\frac{dV%
}{d\varphi }\,=0\,,  \label{5}
\end{equation}%
where dot denotes the derivative with respect to $\varrho .$ If we assume
that the field was initially in the false vacuum state one of the boundary
conditions for $\left( \ref{5}\right) $ is 
\begin{equation}
\varphi \left( \varrho \rightarrow \infty \right) =\varphi _{0}\,.  \label{6}
\end{equation}%
As a second condition Coleman has suggested to use 
\begin{equation}
\dot{\varphi}(\varrho =0)=0  \label{7}
\end{equation}%
to avoid the singularity in the center of the bubble. Equation $\left( \ref%
{5}\right) $ with \textit{these boundary conditions} has an unambiguous
solution $\varphi \left( \varrho \right) $ called the Coleman instanton. The
action for this instanton is given by%
\begin{equation}
S_{I}\,=\,2\pi ^{2}\,\int_{0}^{+\infty }d\varrho \,\varrho ^{3}\,\left( 
\frac{1}{2}\,\dot{\varphi}^{2}\,+\,V(\varphi )\right)  \label{8}
\end{equation}%
and the false vacuum decay rate \textit{per unit time per unit volume} can
be estimated as%
\begin{equation}
\Gamma \simeq \varrho _{0}^{-4}\exp \left( -S_{I}\right) \,,  \label{9a}
\end{equation}%
where $\varrho _{0}$ is the size of the bubble. The pre-exponential factor
is based on dimensional grounds. Calculating the potential energy $%
\left( \ref{3}\right) $ at the moment of Euclidean time $\tau ,$ we infer
that $\mathcal{V}\left( \tau \right) >0$ for $0<\tau <\infty .$ Since the
total energy is normalized to zero this means that in this range of $\tau $
the instanton describes the sub-barrier tunneling in the Euclidean time. As $%
\tau \rightarrow \infty ,$ $\mathcal{V}\rightarrow 0,$ corresponding to the
false vacuum state. The potential energy also vanishes at $\tau =0$ and
hence at this instant\ the bubble emerges from under the barrier in
Minkowski space. To prove that $\mathcal{V}\left( \tau =0\right) =0$ we
first note that for $\varphi \left( \mathbf{x},\tau \right) =\varphi \left(
\varrho \right)$ the expression $\left( \ref{3}\right)$, calculated at $\tau
=0,$ reduces to 
\begin{equation}
\mathcal{V}\left( \tau =0\right) =4\,\pi \int_{0}^{+\infty }d\varrho \,\varrho
^{2}\,\left( \frac{1}{2}\,\dot{\varphi}^{2}\,+\,V(\varphi )\right) \,.
\label{10b}
\end{equation}%
Next we find that the first integral of $\left( \ref{5}\right) $ can be
written as%
\begin{equation}
\frac{1}{2}\dot{\varphi}^{2}-V=\int_{\rho }^{\infty }\frac{3}{\tilde{\varrho}%
}\left( \frac{d\varphi }{d\tilde{\varrho}}\right) ^{2}d\tilde{\varrho}\,,
\label{11a}
\end{equation}%
where the boundary condition $\left( \ref{6}\right) $ has been taken into
account. Finally integrating this equation one gets\footnote{%
The second equality in (\ref{12b}) is the result of integration by parts
taking into account (\ref{6}) and (\ref{7}) as well as requiring that $%
\varphi(\varrho)$ decays faster than $\varrho^{-\frac{1}{2}}$ at $%
\varrho\rightarrow \infty$.} 
\begin{equation}
\int_{0}^{\infty }\left( \frac{1}{2}\,\dot{\varphi}^{2}-V\right) \varrho
^{2}d\varrho =\int_{0}^{\infty }d\left( \varrho ^{3}\right) \left(
\int_{\rho }^{\infty }\frac{1}{\tilde{\varrho}}\left( \frac{d\varphi }{d%
\tilde{\varrho}}\right) ^{2}d\tilde{\varrho}\right) =\int_{0}^{\infty }\dot{%
\varphi}^{2}\varrho ^{2}d\varrho  \label{12b}
\end{equation}%
and hence%
\begin{equation}
\int_{0}^{\infty }\left( \frac{1}{2}\,\dot{\varphi}^{2}+V\right) \varrho
^{2}d\varrho =0  \label{13a}
\end{equation}%
implying that $\mathcal{V}\left( \tau =0\right) $ $\left( \ref{10b}\right) $
really vanishes.

There is a class of potentials for which the results obtained using the
Coleman boundary conditions can either lead to a questionable outcome or
cannot be applied at all, namely,

\textit{a}) for the very steep unbounded potentials (see Fig. \ref{Figure3})
Coleman's instantons lead to nearly instantaneous instability of the false
vacuum, the so called zero size instanton problem,

\textit{b}) for some unbounded potentials, when false vacuum must be
unstable, the Coleman instanton does not exist.

We will show that both problems have common origin and can be resolved by
taking into account inevitable quantum fluctuations which induce the
ultraviolet cutoff scale. In that situation the remedy suggested by Coleman
to avoid a singularity at the origin needs to be and is replaced by an
ultraviolet cutoff scale induced by those very quantum fluctuations. In
turn, this allows us to abandon the very restrictive boundary condition $%
\left( \ref{7}\right) $ and obtain a whole class of new instantons which
would be singular in the absence of this quantum ultraviolet cutoff.

In this paper we will consider an unbounded linear potential and clarify
the role of quantum fluctuations in resolving small instanton problem. The
case of a quartic unbounded potential for which the Coleman instanton does not
exist is analyzed in the companion paper $\cite{MRS2}.$

\section{The model}

Let us consider the classically exactly solvable potential 
\begin{equation}
V\left( \varphi \right) =\left\{ 
\begin{array}{cc}
\lambda _{-}\,\varphi _{0}^{3}\,\varphi +\frac{\lambda _{+}}{4}\,\varphi
_{0}^{4} & \text{for }\varphi <0\,, \\ 
\frac{\lambda _{+}}{4}\,\left( \varphi -\varphi _{0}\right) ^{4} & \text{for 
}\varphi >0\,.%
\end{array}%
\right.  \label{14a}
\end{equation}%
It has a local minimum, corresponding to the false vacuum at $\varphi _{0} $%
. To avoid the problem with one loop quantum corrections, we will assume
that the dimensionless coupling constant $\lambda _{+}$ is always much
smaller than unity. At $\varphi =0$ the $\varphi ^{4}-$potential is matched
to a linear unbounded potential for which the dimensionless constant $%
\lambda _{-}$ can be taken to be large.

\begin{figure}[hbt]
\begin{center}
%\iffigs
\includegraphics[height=60mm]{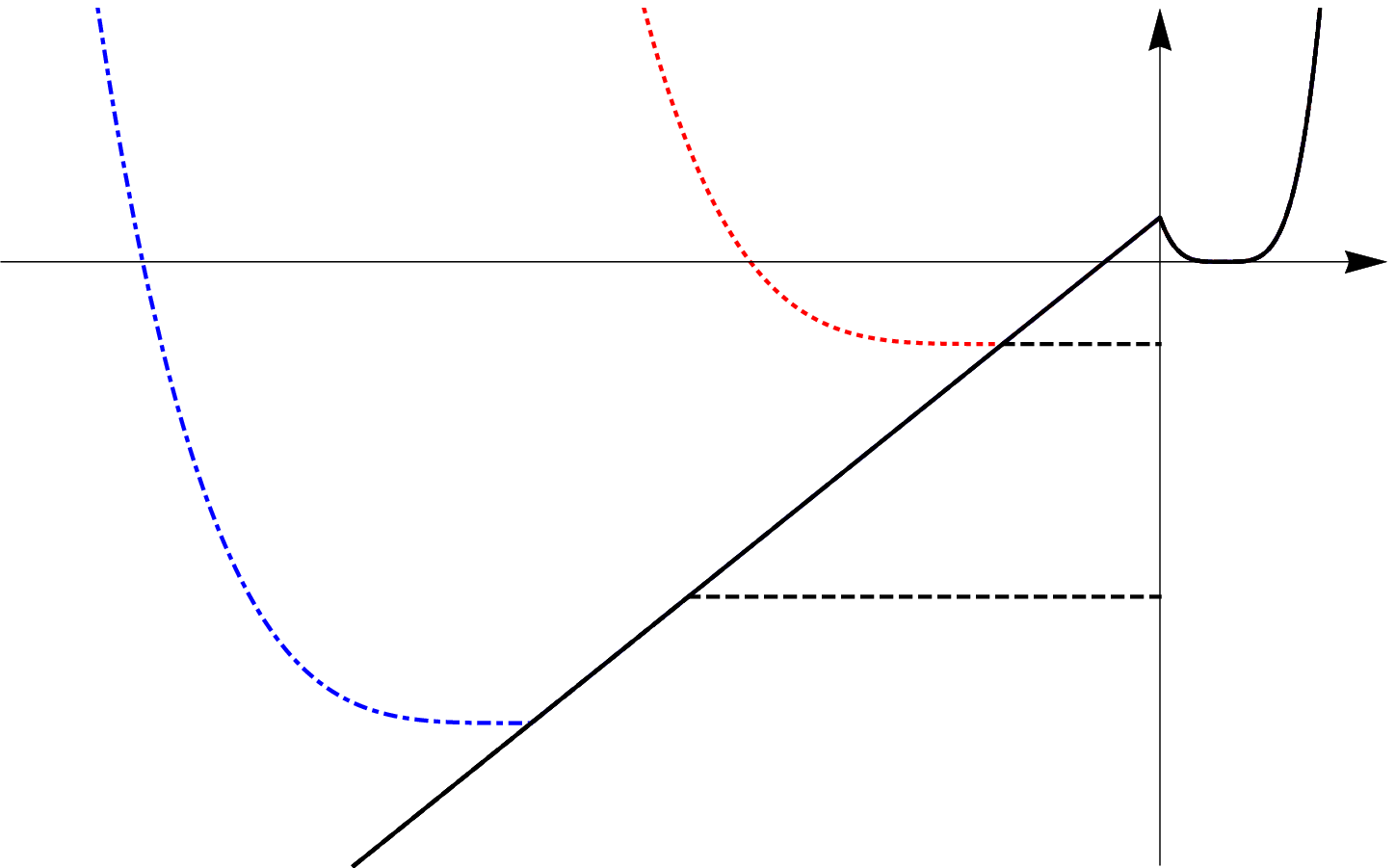} %\else
\end{center}
\par
\vspace*{-4.4cm} \hspace*{11.33cm}$V_{min}$%
\par
\vspace*{1.18cm} \hspace*{11.2cm} $-\frac{4V_{bar}}{\beta }\left( 1-\frac{%
\beta }{4}-{\nu}^{2/3}\right)$%
\par
\vspace*{-3.4cm} \hspace*{10.58cm}$V_{bar}$%
\par
\vspace*{-0.243cm} \hspace*{11.56cm}$\varphi_{0}$%
\par
\vspace*{-1.8cm} \hspace*{3.3cm}$(a)$%
\par
\vspace*{-0.48cm} \hspace*{7.13cm}$(b)$%
\par
\vspace*{-1.0cm} \hspace*{11.1cm}$V$%
\par
\vspace*{1.6cm} \hspace*{12.9cm}$\varphi$%
\par
%%%%%%%%%%
%\fi
\vspace*{4cm} %\iffigs
%\hskip 1cm \unitlength=1.1mm
%\fi
\caption{}
\label{Figure2}
\end{figure}

The results obtained in this paper can also be applied to estimate the
probability of tunneling for some potentials with a second true minimum at
some negative $\varphi _{\min },$ which for $\varphi <0$ can well be
approximated by the linear potential $\left( \ref{14a}\right) $ nearly up to 
$\varphi _{\min }$ (see dotted line in Fig. \ref{Figure2}).

\vspace*{0.9cm}

\section{The Coleman Instanton}

First we find the explicit exact solution for the Coleman instanton and show
how the problem mentioned in the introduction arises in this simple
particular case. For positive $\varphi $ equation $\left( \ref{5}\right) $
takes the following form%
\begin{equation}
\ddot{\varphi}+\frac{3}{\varrho }\dot{\varphi}-\lambda _{+}\left( \varphi
-\varphi _{0}\right) ^{3}=0  \label{15}
\end{equation}%
and its solution with the boundary condition $\left( \ref{6}\right) $ is 
\begin{equation}
\varphi \left( \varrho \right) =\varphi _{0}\frac{\varrho ^{2}-\varrho
_{0}^{2}}{\varrho ^{2}-\varrho _{0}^{2}/(1+\delta )}\,,  \label{16}
\end{equation}%
where%
\begin{equation}
\delta =\frac{4\left( 1+\sqrt{1+\lambda _{+}\,\varphi _{0}^{2}\,\varrho
_{0}^{2}/2}\right) }{\lambda _{+}\,\varphi _{0}^{2}\,\varrho _{0}^{2}}
\label{17}
\end{equation}%
and the constant of integration $\varrho _{0}$ (the size of the bubble) yet
remains to be fixed with the help of the second boundary condition $\left( %
\ref{7}\right) .$ The solution $\left( \ref{16}\right) $ is valid only for $%
\varrho _{0}<\varrho <\infty .$ At $\varrho =\varrho _{0}$ the field $%
\varphi $ vanishes and then becomes negative. For $\varphi <0$ the potential
is linear and equation $\left( \ref{5}\right) $ simplifies to%
\begin{equation}
\ddot{\varphi}+\frac{3}{\varrho }\,\dot{\varphi}-\lambda _{-}\,\varphi
_{0}^{3}=0\,.  \label{18}
\end{equation}%
The solution of this equation, which vanishes at $\varrho _{0}$ and
satisfies $\left( \ref{7}\right) ,$ is 
\begin{equation}
\varphi \left( \varrho \right) =\frac{\lambda _{-}\,\varphi _{0}^{3}}{8}%
\,\left( \varrho ^{2}-\varrho _{0}^{2}\right) \,.  \label{19}
\end{equation}%
The derivative of the field $\varphi $ at $\varrho =\varrho _{0}$ must be
continuous. This allows us to express the size of the bubble in terms of the
parameters $\lambda _{+},\lambda _{-}$ and $\varphi _{0}.$ Equating the
derivatives of solutions $\left( \ref{16}\right) $ and $\left( \ref{19}%
\right) $ at $\varrho _{0}$ we obtain the following equation 
\begin{equation}
1+\frac{1}{\delta }=\frac{\lambda _{-}\,\varphi _{0}^{2}}{8}\,\varrho
_{0}^{2}\,,  \label{20}
\end{equation}%
with $\delta $ given in $\left( \ref{17}\right) .$ Solving this equation for 
$\varrho _{0}^{2}$ one gets 
\begin{equation}
\varrho _{0}^{2}=\frac{8}{\varphi _{0}^{2}\,\lambda _{-}\,\left( 1-\beta
\right) }\,,  \label{21}
\end{equation}%
where we have introduced the parameter 
\begin{equation}
\beta =\frac{\lambda _{+}}{\lambda _{+}+\lambda _{-}}  \label{21a}
\end{equation}%
instead of $\lambda _{+}.$

As one can see from $\left( \ref{16}\right) $ for $\delta \ll 1$ the bubble
has a thin wall. Substituting $\varrho _{0}^{2}$ from $\left( \ref{21}%
\right) $ into equation $\left( \ref{20}\right) $ we find 
\begin{equation}
\delta =\frac{1-\beta }{\beta }  \label{22a}
\end{equation}%
and therefore the thin-wall approximation is valid only if $1-\beta \ll 1,$
that is, for rather flat potential $\left( \lambda _{-}\ll \lambda
_{+}\right) $ at negative $\varphi $. As it follows from $\left( \ref{19}%
\right) $ the value of the scalar field in the center of the bubble is equal
to%
\begin{equation}
\varphi \left( \varrho =0\right) =-\frac{\varphi _{0}}{1-\beta }\,,
\label{23}
\end{equation}%
which corresponds to 
\begin{equation}
V\left( \varphi \left( \varrho =0\right) \right) =-\frac{\lambda
_{-}\,\varphi_{0}^{4}}{1-\beta } \,\left( 1-\frac{\beta }{4}\right) =-\frac{%
4\,V_{bar}}{\beta}\, \left( 1-\frac{\beta }{4}\right) ,  \label{24}
\end{equation}%
where%
\begin{equation}
V_{bar}=\lambda _{+}\,\varphi _{0}^{4}/4  \label{24a}
\end{equation}%
is the height of the barrier. The instanton action can be easily calculated
and is equal to%
\begin{equation}
S_{I}=\frac{8\,\pi ^{2}}{3\,\lambda _{-}\left( 1-\beta \right) ^{3}}\,\left(
2-\beta \right)\, (2-2\beta +\beta ^{2}) \,.  \label{25}
\end{equation}
We note that if one would decide to consider $\lambda_-\rightarrow \infty$, $%
\beta\rightarrow 0$ the value of the scalar field in the center of the
bubble is equal to $\varphi(\varrho=0)\rightarrow -\varphi_0$ and $V\left(
\varphi \left( \varrho =0\right) \right)\rightarrow -\infty$.

\begin{figure}[tbh]
\begin{center}
%\iffigs
\includegraphics[height=60mm]{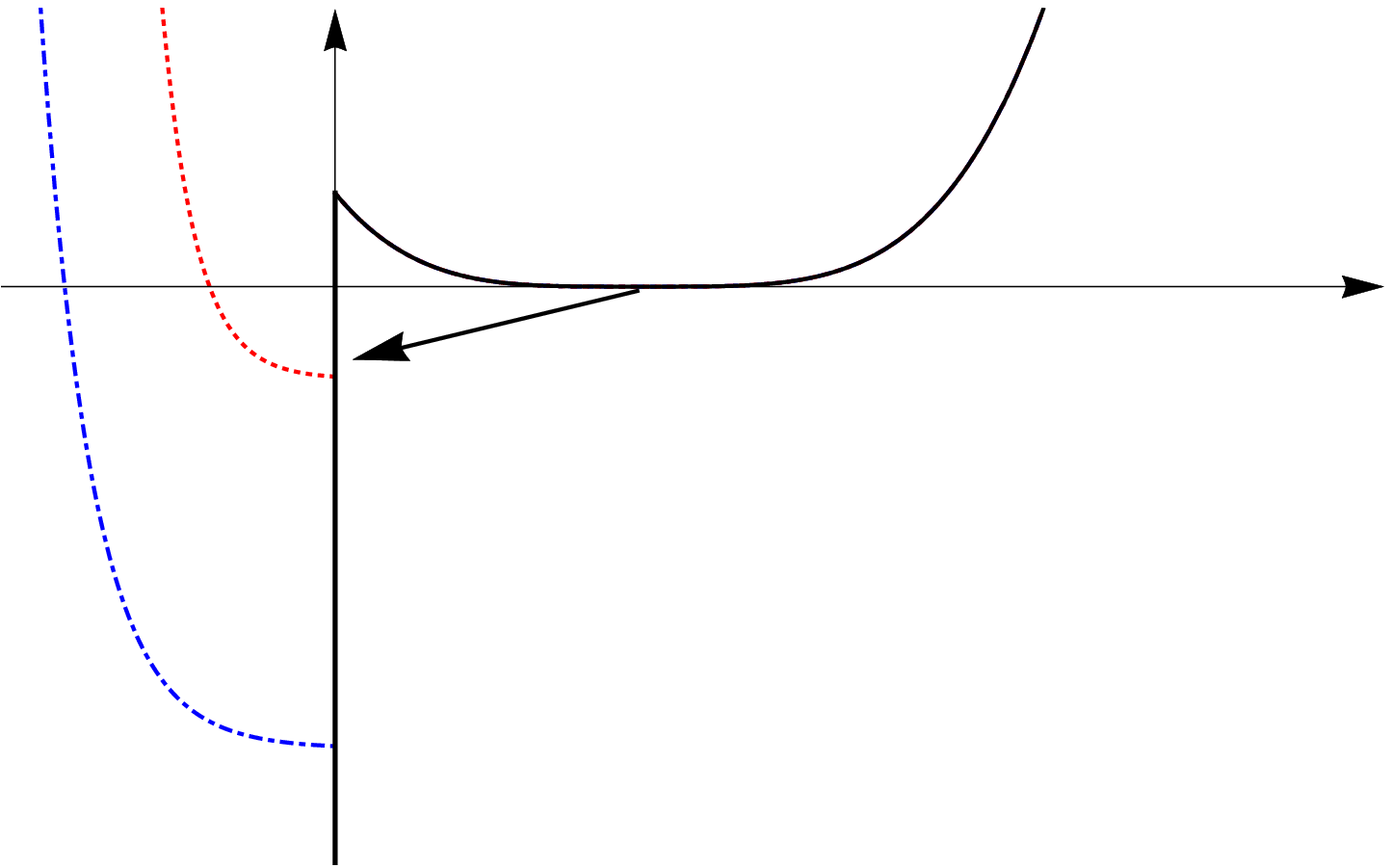} %\else
\end{center}
\par
\vspace*{-4.08cm} \hspace*{5.55cm}$V_{min}$%
\par
\vspace*{1.6cm} \hspace*{5.32cm} -$-\frac{128\,V_{bar}}{\sigma^2\,\lambda_+}$%
\par
\vspace*{-4.1cm} \hspace*{5.58cm}$V_{bar}$%
\par
\vspace*{0.5cm} \hspace*{7.6cm}$\varphi_{0}$%
\par
\vspace*{-2.5cm} \hspace*{2.88cm}$(a)$%
\par
\vspace*{-0.46cm} \hspace*{3.78cm}$(b)$%
\par
\vspace*{-0.9cm} \hspace*{5.4cm}$V$%
\par
\vspace*{1.7cm} \hspace*{13.0cm}$\varphi$%
\par
%%%%%%%%%%
%\fi
\vspace*{4.0cm} %\iffigs
%\hskip 1cm \unitlength=1.1mm
%\fi
\caption{{}}
\label{Figure3}
\end{figure}

\textit{Instantaneous vacuum decay via small bubbles.} For a very steep
unbounded potential shown in Fig. \ref{Figure3} and obtained by taking the
limit $\lambda _{-}\rightarrow \infty ,$ the bubble size given in $\left( %
\ref{21}\right) $ shrinks to zero as $\varrho _{0}\propto \lambda
_{-}^{-1/2} $ and the action $\left( \ref{25}\right) $ vanishes$.$ The false
vacuum decay rate $\left( \ref{9a}\right) $ becomes infinite irrespective
how high is the potential barrier $V_{bar}$. This instantaneous false vacuum
decay happens via infinitely small bubbles, which according to $\left( \ref%
{24}\right) $ emerge with infinitely large negative potential in the center.
Such a conclusion does not sound physically acceptable. Moreover, even for
finite but large enough $\lambda _{-}$ the problem of small instantons still
remains because they lead to unexpectedly efficient decay with the rate
practically independent on the shape of the potential at positive $\varphi $%
. The situation starts to look even more strange if one assumes that the
potential in Fig. \ref{Figure3} gets a second rather sharp minimum (see
dotted line). Then, irrespective of the depth of this second minimum, it
looks like the Coleman instanton ceases to exist and, thus, nearly
irrelevant modification of the potential abruptly changes the decay rate
from infinity to zero.

\textbf{\ }

Below we will show how these problems are resolved if we take into account
quantum fluctuations.

\section{Quantum Fluctuations}

The instanton is a \textit{classical} solution and it is reliable only if
both the field and its derivative exceed the level of the minimal quantum
fluctuations. The bubble emerges in Minkowski space at $\tau =0$. At this
instant the distance to the center of the bubble is equal to $\varrho $ and
the \textquotedblleft typical\textquotedblright\ amplitude of the quantum
fluctuations in corresponding scales is about (see for example, \cite%
{Mukhanov2}): 
\begin{equation}
\left\vert \delta \varphi _{q}\right\vert \simeq \frac{\sigma}{\varrho }\,,
\label{27}
\end{equation}%
and, respectively, its time derivative is of order%
\begin{equation}
\left\vert \delta \dot{\varphi}_{q}\right\vert \simeq \frac{\sigma }{\varrho
^{2}}\,,  \label{28}
\end{equation}%
where $\sigma $ is the numerical coefficient of order unity (we set $\hbar=1$%
). The quantum fluctuations grow very fast as $\varrho \rightarrow 0$ and,
hence, Coleman's boundary condition $\dot{\varphi}_{\varrho =0}=0,$
formulated in the deep ultraviolet limit, must be re-analyzed. The potential 
$V\left( \varphi \right) $ can be arbitrarily shifted along the $\varphi$%
--axis. Therefore, to estimate the magnitude of quantum fluctuations it is
more appropriate to consider either the typical change of the classical
field $\Delta \varphi \simeq \dot{\varphi}\times \varrho $ in scales $%
\varrho $ or it's time derivative. In both cases we obtain (up to a
numerical coefficient) the same result and to be concrete we will use $%
\left( \ref{28}\right) $ to determine when quantum fluctuations become
relevant. Equating the derivative of the classical solution $\left( \ref{19}%
\right) $ to the amplitude of quantum fluctuations $\left( \ref{28}\right) $
we find the following ultraviolet cutoff scale%
\begin{equation}
\varrho _{uv}=\left( \frac{4\,\sigma }{\lambda _{-}\,\varphi _{0}^{3}}%
\right) ^{1/3}=\left( \frac{\sigma ^{2}}{32}\right) ^{1/6}\lambda
_{-}^{1/6}\,\left( 1-\beta \right) ^{1/2}\varrho _{0}\,,  \label{29}
\end{equation}%
where $\beta $ is defined in $\left( \ref{21a}\right) $ and $\varrho _{0}$
is given in $\left( \ref{21}\right) .$ The classical solution with the
Coleman boundary condition $\left( \ref{7}\right) $ is obviously valid only
for $\varrho >\varrho _{uv}$ and it is completely spoiled by quantum
fluctuations on scales smaller than $\varrho _{uv}$. Let us notice that for
the very steep potentials with $\lambda _{-}\gg 1$ the ultraviolet cutoff
scale exceeds the size of the bubble and therefore the instantons which lead
to the large decay rate cannot be trusted anymore.

The ultraviolet cutoff $\varrho _{uv}$ resulted from the fact that the
quantum fluctuations are characterized by the field derivative of order $%
1/\varrho ^{2}$ for all values of $\varrho $ while the contribution of the
classical solution at small $\varrho $ decreases as $\varrho $. Thus, for
small enough $\varrho $ the quantum fluctuations dominate. For the large
values of $\varrho $ the quantum fluctuations continue to exhibit a $%
1/\varrho ^{2}$ decay while the classical solution decreases as $1/\varrho
^{3}$. Thus, the quantum fluctuations take over both in the deep
ultra-violet and the large infrared scales carving a range of values of $%
\varrho $ for which the classical solutions can be valid.

Next we estimate the infrared cutoff scale above which (in length) quantum
fluctuations dominate. To find this scale we have to equate the derivative
of solution $\left( \ref{16}\right) ,$ valid for $\varrho >\varrho _{0},$ to 
$\left( \ref{28}\right) .$ As a result one gets%
\begin{equation}
\varrho _{ir}\simeq \frac{2}{\sigma }\frac{\delta }{1+\delta }\,\varphi
_{0}\,\varrho _{0}^{2}\simeq \frac{4\sqrt{2}}{\sigma }\lambda
_{-}^{-1/2}\left( 1-\beta \right) ^{1/2}\varrho _{0}\,.  \label{30}
\end{equation}%
To simplify the result we have assumed that $\varrho _{ir}\gg \varrho _{0}.$
As one can check a posteriori this assumption is really valid when both $%
\lambda _{+}$ and $\lambda _{-}$ are much smaller than unity.

For $\lambda _{-}>1$ we have $\varrho _{ir}<\varrho _{0}$ and moreover $%
\varrho _{uv}>\varrho _{ir}$. Thus, the range 
\begin{equation}
\varrho _{uv}<\varrho <\varrho _{ir}  \label{31}
\end{equation}%
where the classical instanton solution is supposed to be valid completely
disappears. This tells us that for $\lambda _{-}>1$ the classical instanton
with the Coleman boundary condition can not be applied.

Concluding this section we would like to stress that both the \textit{%
ultraviolet and infrared cutoff scales} \textit{are entirely determined by
the parameters characterizing the corresponding classical solution.}

\section{New Instantons}

\label{section5}

The existence of the cutoff scales allows us to obtain a new class of $%
O\left( 4\right) $ instantons, which otherwise would be singular and having
an infinite action would not contribute to the decay rate. Let us first
consider the new solutions emerging thanks to the existence of the
ultraviolet cutoff scale. Later on we will estimate how the infrared cutoff
corrections influence these new solutions. Thus, we will assume that for $%
\varrho >\varrho _{0}$ solution $\left( \ref{16}\right) -\left( \ref{17}%
\right) $ is still valid, but abandon the boundary condition $\dot{\varphi}%
_{\varrho =0}=0.$ Then the most general solution of equation $\left( \ref{18}%
\right) $, which vanishes at $\varrho _{0}=0$ and valid for $\varphi <0$ is%
\begin{equation}
\varphi \left( \varrho \right) =\frac{\lambda_{-}\,\varphi _{0}^{3}}{8}\left(
\varrho ^{2}-\varrho _{0}^{2}\right) \left( 1-\frac{\varrho _{1}^{2}}{%
\varrho ^{2}}\right) ,  \label{32}
\end{equation}%
where the constant of integration $\varrho _{1}<\varrho _{0}$ parametrizes
our new instantons. These instantons are singular at $\varrho \rightarrow 0.$
However, as we have shown above the solution $\left( \ref{32}\right) $ can
only be trusted for $\varrho >\varrho _{uv}$. This allows us to avoid a
singularity in the classical solution. When $\varrho _{1}\neq 0$ there is a
bounce at 
\begin{equation}
\varrho _{b}=\sqrt{\varrho_{0}\,\varrho _{1}}\,,  \label{33}
\end{equation}%
where $\dot{\varphi}(\varrho _{b})=0,$ and after that the field $\varphi $
must go back and vanish at $\varrho _{1},$ which is smaller than $\varrho
_{b}.$ However, before this happens the quantum fluctuations become relevant
at $\varrho _{uv}>\varrho _{b}$ and as we will show the classical bubble
with a quantum core emerges in Minkowski space$.$ To calculate $\varrho
_{uv} $ we equate the derivative of $\left( \ref{32}\right) $ to $\left( \ref%
{28}\right) $ and obtain the following equation $\qquad $
\begin{equation}
1-\frac{\varrho _{b}^{4}}{\varrho _{uv}^{4}} =\frac{4\,
\sigma}{\lambda_{-}\varphi _{0}^{3}}\,\frac{1}{\varrho_{uv}^{3}}\,,  \label{34}
\end{equation}%
which can be solved exactly for $\varrho _{uv}$. Because we will not need
the explicit solution for $\varrho _{uv}$ we skip it here, but instead let
us \ rewrite equation $\left( \ref{34}\right) $ in more convenient form as%
\begin{equation}
\frac{\varrho _{uv}^{2}-\varrho _{b}^{2}}{\varrho _{uv}^{2}}=\nu \left( 
\frac{\bar{\varrho}}{\varrho _{uv}}\right) ^{3},  \label{35}
\end{equation}%
where%
\begin{equation}
\bar{\varrho}^{2}\equiv \frac{8}{\varphi_{0}^{2}\,\lambda _{-}\,\left( 1-\beta
\right) }  \label{35a}
\end{equation}%
and 
\begin{equation}
\nu \equiv \frac{\sigma }{4\,\sqrt{2}\,\kappa\, \left( \varrho _{uv}\right) }%
\,\lambda_{-}^{1/2}\,\left( 1-\beta \right) ^{3/2}.  \label{36}
\end{equation}%
Here, $\kappa \left( \varrho _{uv}\right) =1+\varrho _{b}^{2}/\varrho_{uv}^{2} $ and since $\varrho _{b}^{2}/\varrho _{uv}^{2}<1$ it
implies that $1<\kappa \left( \varrho _{uv}\right) <2.$ The scale $\bar{%
\varrho}$ entering in $\left( \ref{35}\right) $ is equal to the bubble size $%
\varrho _{0}$ (see (\ref{21})) only for $\varrho _{1}=0,$ corresponding to
the Coleman instanton. To determine $\varrho _{0}$ in general case we require
that $\dot{\varphi}(\varrho )$ is continuous at $\varrho _{0}$ and equating
the derivatives of $\left( \ref{16}\right) $ and $\left( \ref{32}\right) $
at this point we obtain the following equation 
\begin{equation}
1+\frac{1}{\delta }=\frac{\lambda _{-}\,\varphi_{0}^{2}}{8}\,(\varrho_{0}^{2}-\varrho _{1}^{2})\,,  \label{38}
\end{equation}%
where $\delta $ is given in $\left( \ref{17}\right) .$ Solving this equation
for $\varrho _{0}^{2}$ one gets 
\begin{equation}
\varrho _{0}^{2}=\frac{1}{2}\left( 1+\sqrt{1+4\,\beta \,\frac{\varrho _{1}^{2}}{%
\bar{\varrho}^{2}}}~\right) \bar{\varrho}^{2}+\varrho_{1}^{2}\,.  \label{39}
\end{equation}

Thus we have found the new instantons with finite action, which are
parametrized by $\varrho _{1}$. The main contribution to the decay rate
comes from those instantons which have the minimal action while describing
the required transition. Let us stress that $\varrho _{1}^{2}$ must be
positive because otherwise the growing mode $\varphi \propto 1/\varrho ^{2}$
would be increasing faster than the quantum fluctuations as $\varrho
\rightarrow 0$ and we would end up at a singularity, where $\varphi
\rightarrow -\infty .$ For positive $\varrho _{1}^{2}$ the classical field $%
\varphi $ bounces at $\varrho _{b}$ and evolves towards the positive values.
However, as we have noticed above, before the bounce is reached the
classical solution fails at $\varrho _{uv}>\varrho _{b}$ and the bubble
emerges from under the barrier materializing in Minkowski space.

To simplify the formulae it is convenient to parametrize the instantons by
the dimensionless parameter $\chi \left( \varrho _{1}\right) ,$ related to $%
\varrho _{1}^{2}$ as%
\begin{equation}
\varrho _{1}^{2}=\chi \left( 1+\beta \,\chi \right) \bar{\varrho}^{2}\,,
\label{40}
\end{equation}%
where $\bar{\varrho}^{2}$ is defined in $\left( \ref{35a}\right) $. Then, as
it follows from $\left( \ref{39}\right) $ 
\begin{equation}
\varrho _{0}^{2}=\left( 1+\chi \right) \left( 1+\beta\, \chi \right) \bar{%
\varrho}^{2}  \label{42}
\end{equation}%
and%
\begin{equation}
\varrho _{b}^{2}=\varrho _{0}\,\varrho _{1}=\left( 1+\beta\, \chi\, \right) \sqrt{%
\chi \,(1+\chi )}\,\bar{\varrho}^{2}\,.  \label{43}
\end{equation}%
The expression $\left( \ref{32}\right) $ for $\varphi \left( \varrho \right) 
$ can be rewritten in the following more convenient form: 
\begin{equation}
\varphi \left( \varrho \right) =-\frac{\varphi _{0}}{1-\beta }\left( \frac{%
\left( \varrho _{0}-\varrho _{1}\right) ^{2}}{\bar{\varrho}^{2}}-\frac{%
\left( \varrho ^{2}-\varrho _{b}^{2}\right) ^{2}}{\bar{\varrho}^{2}\varrho
^{2}}\right)  \label{44a}
\end{equation}%
and using the formulae above as well as equation $\left( \ref{35}\right) $
we obtain%
\begin{equation}
\varphi _{uv}\equiv \varphi \left( \varrho _{uv}\right) =-\frac{\varphi_{0}%
}{1-\beta }\left( \frac{1+\beta \,\chi }{\left( \sqrt{1+\chi }+\sqrt{\chi }%
\right) ^{2}}-\nu ^{2}\left( \frac{\bar{\varrho}}{\varrho_{uv}}\right)^{4}\right) \,.  \label{45}
\end{equation}%
The value of the potential at $\varphi _{uv}$ is 
\begin{equation}
V_{uv}\left( \chi \right) =V\left( \varphi _{uv}\right) =-\frac{4V_{bar}}{%
\beta }\left( \frac{1+\beta\, \chi }{\left( \sqrt{1+\chi }+\sqrt{\chi }\right)^{2}}
-\frac{\beta }{4}-\nu^{2}\,\left( \frac{\bar{\varrho}}{\varrho _{uv}}%
\right)^{4}\right) \,.  \label{46}
\end{equation}%
Assuming that $\lambda _{-}\ll 1$ we find that when $\chi $ varies from zero
to infinity $V_{uv}$ changes within the range 
\begin{equation}
-\frac{4V_{bar}}{\beta }\left( 1-\frac{\beta }{4}-\nu ^{2/3}\right)
<V_{uv}\left( \chi \right) <0\,,  \label{46a}
\end{equation}%
vanishing at $\chi \rightarrow \infty $. The last term in the brackets is
due to the quantum corrections and it is much smaller than unity for $%
\lambda _{-}\ll 1$ (the case of $\lambda _{-}>1$ will be considered
separately).

Thus, we conclude that as a result of quantum regularization there appears a
whole class of new nonsingular instantons which all contribute to the decay
of the false vacuum. Depending on $\chi $, which parametrizes these
instantons, the value of the potential in the central part of the bubble
dominated by quantum fluctuations, after subtracting the energy of zero
point fluctuations, is equal to $V_{uv}$ and takes its value within the
interval $\left( \ref{46a}\right) .$ If there exists a second true vacuum
with $V_{\min }$ in the range $\left( \ref{46a}\right) $ (see curve (b) in
Fig. \ref{Figure2}), we expect that the dominant contribution to the false
vacuum decay is given by the instanton with $\chi $ determined by equation $%
V_{uv}\left( \chi \right) \simeq V_{\min }.$

To verify that the new instanton really emerges from under the potential
barrier in Minkowski space at $\tau =0$ we have to check that%
\begin{equation}
\mathcal{V}\left( \tau =0\right) =4\,\pi \int_{\varrho _{uv}}^{+\infty
}d\varrho \,\varrho ^{2}\,\left( \frac{1}{2}\,\dot{\varphi}%
^{2}\,+\,V(\varphi )\right) +\frac{4\,\pi }{3}\,\varrho_{uv}^{3}V_{uv},
\label{47}
\end{equation}%
vanishes up to a possible quantum correction not exceeding the contribution
of a single quantum. The second term in $\left( \ref{47}\right) $ accounts
for the shifted energy inside the bubble quantum core, which remains after
normalizing the energy of quantum fluctuations to zero in the false vacuum
state. Substituting solutions $\left( \ref{16}\right) $ and $\left( \ref{32}%
\right) $ in $\left( \ref{47}\right) $ we obtain%
\begin{eqnarray}
&& 4\,\pi \int_{\varrho _{uv}}^{+\infty }d\varrho \,\varrho ^{2}\,\left( \frac{1%
}{2}\,\dot{\varphi}^{2}\,+\,V(\varphi )\right)  \notag \\
&=&\frac{\pi\, \lambda_{-}^{2}\,\varphi _{0}^{6}}{24}\left[ \frac{\left(
\varrho _{uv}^{4}-\varrho _{b}^{4}\right) ^{2}}{\varrho _{uv}^{3}}-4\varrho
_{uv}\left( \left( \varrho _{0}^{2}-\varrho _{uv}^{2}\right) \left( \varrho
_{1}^{2}-\varrho _{uv}^{2}\right) +\frac{2\lambda _{+}}{\lambda
_{-}^{2}\varphi _{0}^{2}}\varrho _{uv}^{2}\right) \right]  \notag \\
&=&\frac{4\,\pi }{3}\varrho_{uv}^{3}\left( \frac{1}{2}\left( \dot{\varphi}%
\left( \varrho _{uv}\right) \right) ^{2}-V\left( \varphi \left( \varrho
_{uv}\right) \right) \right) ,  \label{48}
\end{eqnarray}%
and hence%
\begin{equation}
\mathcal{V}\left( \tau =0\right) =\frac{2\,\pi }{3}\varrho _{uv}^{3}\left( 
\dot{\varphi}\left( \varrho _{uv}\right) \right) ^{2}=\frac{2\,\pi\, \sigma ^{2}%
}{3\,\varrho _{uv}}\,,  \label{49}
\end{equation}%
which corresponds to one quantum of energy in scales $\varrho _{uv}.$ Thus,
the energy balance is satisfied within an accuracy dictated by the
time-energy uncertainty relation and the bubble emerges from under the
barrier.

To determine the decay rate we calculate the action%
\begin{equation}
S_{I}\,=\,2\pi ^{2}\,\int_{\varrho _{uv}}^{+\infty }d\varrho \,
\varrho^{3}\,\left( \frac{1}{2}\,\dot{\varphi}^{2}\,+\,V(\varphi )\right) +%
\frac{\pi ^{2}}{2}\varrho _{uv}^{4}V_{uv}\,,  \label{50}
\end{equation}%
where the last term accounts for the contribution of the bubble quantum
core. The result is%
\begin{eqnarray}
S_{I}\, &=&\frac{\pi \,\varphi_{0}^{2}}{96}\left[\lambda_{-}^{2}\,\varphi_{0}^{4}\,
\left( \varrho_{uv}^{6}+\frac{3\varrho_{b}^{8}}{\varrho _{uv}^{2}}%
-\varrho_{0}^{6}-\frac{3\,\varrho_{b}^{8}}{\varrho_{0}^{2}}\right) +\frac{%
32\,\left( \lambda _{-}\varphi _{0}^{2}\left( \varrho _{0}^{2}-\varrho
_{1}^{2}\right) -8\right) }{\lambda _{+}\varphi _{0}^{2}}\right.  \notag \\
&&\left. +\left( 16\,\lambda _{-}\,\varphi _{0}^{2}\left( \left( 1+\frac{%
3\,\lambda _{+}}{4\lambda_{-}}\right) \varrho _{0}^{2}
-\varrho_{1}^{2}\right) \right) \varrho_{0}^{2}+32\,\varrho_{0}^{2}\right]\,.
\label{51}
\end{eqnarray}%
Substituting here the expression for $\varrho _{0},\varrho _{1}$ and $%
\varrho _{b}$ from $\left( \ref{40}\right) -\left( \ref{43}\right) $ and
using equation $\left( \ref{35}\right) $ for $\varrho _{uv}$ this action can
be represented in the following form:%
\begin{eqnarray}
S_I &=&\frac{8\pi ^{2}}{3\lambda _{-}\left( 1-\beta \right) ^{3}}\Bigg[ %
\left( 2-\beta \right) (2-2\beta +\beta ^{2})+\frac{4\,\chi^{3/2}\left(
4+3\,\chi \right) \,\left( 1+\beta \,\chi \right) ^{3}}{ 2\,\left( 1+\chi
\right) ^{3/2}+\chi^{1/2}\left( 3+2\,\chi \right) }  \notag \\
&-&\beta\, \chi^{2}\left( 1+2\,\left( 1+\beta \,\chi \right) +3\left( 1+\beta\,
\chi \right)^{2}\right) \Bigg] +\frac{\pi^{2}\,\sigma ^{2}}{6}\left(
1+2\,\left( \frac{\varrho_{b}^{2}}{\varrho_{uv}^{2}+\varrho_{b}^{2}}\right)
^{2}\right) \,.  \label{52}
\end{eqnarray}%
The last term here is always of order unity and can be neglected.

\textbf{Remarks on infrared cutoff}\textit{. } Considering instantons we
have ignored the infrared cutoff. Let us estimate how the results above are
changed if we take it into account. For $\chi \neq 0$ the expression $\left( %
\ref{30}\right) $ for $\varrho _{ir}$ is modified as 
\begin{equation}
\varrho _{ir}\simeq \frac{4\sqrt{2}}{\sigma }\lambda _{-}^{-1/2}\left(
1-\beta \right) ^{1/2}\left( \frac{1+\chi }{1+\beta \chi }\right)
^{1/2}\varrho _{0}\,.  \label{53}
\end{equation}%
If both $\lambda _{+},\lambda _{-}\ll 1$ the bubble size $\varrho _{0}$ is
always much smaller than $\varrho _{ir}$ and the infrared effects do not
influence much the instantons for any $\chi $. However, if $\lambda _{-}\gg
1,$ it follows from $\left( \ref{53}\right) $ that only if 
\begin{equation}
\chi \gg \chi _{\min }=4\,\nu ^{2}\,,  \label{54}
\end{equation}%
where $\nu $ is defined in $\left( \ref{36}\right) ,$ we have $\varrho
_{ir}\gg \varrho _{0},$ otherwise $\varrho _{ir}$ can becomes even smaller
than $\varrho _{uv}$ and, hence, the classical solutions with $\chi <\chi
_{\min }$ do not make sense. Thus, for the steep potentials with $\lambda
_{-}\gg 1$ the value of $\chi $ must always exceed $\chi _{\min }\approx
\lambda _{-}.$

Calculating the corrections to the potential energy $\left( \ref{49}\right) $
and action (\ref{50}) due to the infrared cutoff, we find that%
\begin{equation}
\Delta \mathcal{V}_{ir}=-4\pi \int_{\varrho _{ir}}^{+\infty }d\varrho \,
\varrho^{2}\,\left( \frac{1}{2}\,\dot{\varphi}^{2}\,+\,V(\varphi )\right) =%
\frac{2\,\pi \,\sigma ^{2}}{3}\left( 1+\frac{\sigma^{2}\,\lambda_{+}}{60}\right) 
\frac{1}{\varrho _{ir}}  \label{55}
\end{equation}%
and 
\begin{equation}
\Delta S_{I}=-\frac{\pi ^{2}\,\sigma^{2}}{32}\left( 1+\frac{\sigma^{2}\,\lambda _{+}}{4}\right) \,,  \label{56}
\end{equation}%
respectively. If we would decide to normalize $V\left( \varphi \left(
\varrho _{ir}\right) \right) $ to zero we would have to subtract from the
action 
\begin{equation}
\Delta S=-\frac{\pi ^{2}\,\sigma^{4}\,\lambda_{+}}{128} \,.  \label{57}
\end{equation}%
Thus, as it follows from $\left( \ref{55}\right) -\left( \ref{57}\right) $
all infrared corrections are at the level of one quantum with the frequency
corresponding to the infrared cutoff scale. Hence, except the important
bound $\left( \ref{54}\right)$ which has to be respected for $\lambda
_{-}\gg 1$, the quantum infrared corrections can be ignored.

\section{Implications}

We will now apply the results obtained above in several limiting cases. We
consider separately the instantons with $\chi \ll 1$ and $\chi \gg 1.$ The
action $\left( \ref{52}\right) $ simplifies to 
\begin{equation}
S_{I}\left( \chi \right) =\frac{8\,\pi ^{2}}{3\,\lambda _{-}\left( 1-\beta
\right) ^{3}}\left[ \left( 2-\beta \right) (2-2\,\beta +\beta^{2})
+8\,\chi^{3/2}+O\left(\chi ^{2}\right) \right]  \label{57a}
\end{equation}%
for $\chi \ll 1$ and up to corrections of order $\chi ^{3/2}$ coincides with
the Coleman action, while for $\chi \gg 1$ it becomes%
\begin{equation}
S_{I}=\frac{8\,\pi^{2}\,\chi }{\lambda_{-}\left(1-\beta \right) ^{3}}\,
\left[\frac{1}{3}\left( 1-\frac{\beta}{2}\right) \left(\beta \,\chi \right)
^{2}+\left( 1-\frac{\beta}{4}\right)^{2}\,\beta\, \chi +\left( 1-\frac{\beta }{%
4}\right) \left( 1-\frac{\beta }{4}+\frac{\beta ^{2}}{8}\right) +O\left( 
\frac{1}{\chi }\right) \right] .~  \label{57b}
\end{equation}%
Let us recall that in order to ignore the quantum loop corrections $\lambda
_{+}$ must always be smaller than unity, while $\lambda _{-}$ can be
 large and therefore we investigate the potentials with $\lambda
_{-}\ll 1$ and $\lambda _{-}\gg 1$ separately.

\textbf{A)} For $\lambda _{-}\ll 1$ the parameter $\nu $ defined in $\left( %
\ref{36}\right) $ is much smaller than unity and thus from $\left( \ref{46}%
\right) $ one obtains 
\begin{equation}
V_{uv}\left( \chi \right) =-\frac{4\,V_{bar}}{\beta }\left( 1-\frac{\beta }{4}%
-2\,\sqrt{\chi }-\nu^{2/3}+O\left( \chi \right) \right)  \label{58}
\end{equation}%
for $\chi \ll 1.$ These instantons with the action $\left( \ref{57a}\right) $
give the main contribution to the decay rate for unbounded potential and for
the potentials with the second true minimum of depth $V_{\min }<V_{uv}\left(
\chi =0\right) .$ They are different from the Coleman instanton ($\chi =0)$ only
by higher order corrections. As it follows from $\left( \ref{35}\right) $
the size of quantum core%
\begin{equation}
\varrho _{uv}\simeq \left( \frac{\sigma ^{2}}{32}\right) ^{1/6}\,
\lambda_{-}^{1/6}\,\left( 1-\beta \right)^{1/2}\varrho _{0}  \label{60}
\end{equation}%
is much smaller than the bubble size 
\begin{equation}
\varrho _{0}=\left( 1+\frac{1+\beta}{2}\,\chi +O\left( \chi ^{2}\right)
\right) \bar{\varrho}\,.  \label{60a}
\end{equation}%
To calculate the contribution of instantons with the value of the potential
in the quantum core%
\begin{equation}
V_{uv}\left( \chi _{\varepsilon }\right) =\varepsilon \,V_{uv}\,
\left(\chi=0\right) \simeq -\frac{4\,\varepsilon \,V_{bar}}{\beta }\left( 1-\frac{\beta }{4%
}\right) \,,  \label{60b}
\end{equation}%
where $\varepsilon \ll 1,$ to the decay rate we have to consider solutions
with $\chi \gg 1.$ In this case the expression $\left( \ref{46}\right) $
simplifies to%
\begin{equation}
V_{uv}\left( \chi \right) =-\frac{V_{bar}}{\beta }\left( \left( 1-\frac{%
\beta }{2}\right) \frac{1}{\chi }+O\left( \frac{1}{\chi ^{2}}\right) \right)
,  \label{63}
\end{equation}%
and as it follows from $\left( \ref{60b}\right) $ 
\begin{equation}
\chi _{\varepsilon }\simeq \frac{1}{2}\left( \frac{2-\beta }{4-\beta }%
\right) \frac{1}{\varepsilon }\gg 1\,.  \label{64}
\end{equation}%
The expressions for $\varrho _{0}^{2}$ becomes%
\begin{equation}
\varrho _{0}^{2}\left(\chi_{\varepsilon }\right) \simeq \frac{4}{\left(
1-\beta \right) }\left( \frac{2-\beta }{4-\beta }\right) ^{2}\frac{1}{%
\varepsilon \lambda _{-}\varphi _{0}^{2}}\left( \left( \frac{4-\beta }{%
2-\beta }\right) +\frac{1}{2}\,\frac{\beta }{\varepsilon }\right) ,  \label{65}
\end{equation}%
and the action $\left( \ref{57b}\right) $ is equal to%
\begin{equation}
S_{\varepsilon }=\frac{\pi ^{2}\left( 2-\beta \right) }{\varepsilon \,\lambda
_{-}\left( 1-\beta \right) ^{3}}\left[ \frac{1}{6}\left( \frac{2-\beta }{%
4-\beta }\right) ^{3}\frac{\beta^{2}}{\varepsilon^{2}}+\frac{\left(
2-\beta \right) }{8}\frac{\beta }{\varepsilon }+\left( 1-\frac{\beta }{4}+%
\frac{\beta ^{2}}{8}\right) +O\left( \varepsilon \right) \right] \,.
\label{67}
\end{equation}%
The contribution of these $\chi _{\varepsilon }$-instantons to the decay
rate is given by%
\begin{equation}
\Gamma \sim \varrho _{0}^{-4}\left( \chi _{\varepsilon }\right) \exp \left(
-S_{\varepsilon }\right) \,.  \label{67a}
\end{equation}%
For the unbounded potential the instantons with $\chi \ll 1$ give the main
contribution to the overall decay rate. However, even in this case the whole
spectrum of instantons is present when we consider the false vacuum decay.
The formula $\left( \ref{67a}\right) $ can also be applied to estimate the
leading contribution to the decay rate for a broad class of potentials with
two minima in the case when the potential is well approximated by the linear
potential nearly up to the second true minimum of depth
\begin{equation}
V_{\min }\simeq -\frac{4\,\varepsilon \,V_{bar}}{\beta }\left( 1-\frac{\beta 
}{4}-\nu ^{2/3}\right)
\end{equation}%
(see, for example, curve (b) in Fig. \ref{Figure2}).

The expression $\left( \ref{17}\right) $ for the parameter $\delta ,$
characterizing the thickness of the bubble wall for $\chi \neq 0,$ is
modified as 
\begin{equation}
\delta =\frac{1-\beta }{\beta \left( 1+\chi \right) },  \label{70a}
\end{equation}%
and hence the bubble has a thin wall $\delta \ll 1$ for all $\chi $ if $%
\lambda _{-}\ll \lambda _{+}$ $\left( 1-\beta \ll 1\right) .$ In the case $%
\lambda _{+}\ll \lambda _{-}<1$ only bubbles with $\chi _{\varepsilon }\sim
\varepsilon ^{-1}\gg \lambda _{-}/\lambda _{+}$ have a thin wall.

When we have two nearly degenerate minima, that is $\varepsilon \rightarrow
0 $ ($\varepsilon \ll \beta $), the probability of the vacuum decay, for
example, for $\beta \ll 1$ vanishes as 
\begin{equation}
\Gamma \sim4\, \lambda _{+}^{2}\left( \varphi _{0}\frac{\varepsilon }{\beta }%
\right) ^{4}\exp \left[ -\frac{\pi ^{2}}{24\lambda _{+}}\left( \frac{\beta }{%
\varepsilon }\right) ^{3}\right]  \label{70b}
\end{equation}%
in complete agreement with our expectations.

\textbf{B) }$\lambda _{-}\gg 1$ (\textit{zero size instanton problem}).
Finally let us consider the most interesting case of a very steep unbounded
potential (see Fig. \ref{Figure3} where $\lambda _{-}\rightarrow \infty $).
Since $\lambda _{+}$ is always smaller than unity $\beta \ll 1$ and the
parameter $\nu ^{2}$ in $\left( \ref{36}\right) $ is 
\begin{equation}
\nu ^{2}\simeq \frac{\sigma ^{2}}{128}\,\lambda _{-}\gg 1\,.  \label{77}
\end{equation}%
As one can find by inspecting $\left( \ref{45}\right) $ and $\left( \ref{46}%
\right) ,$ in this case the value of $\chi $ must be larger than unity
because otherwise the quantum fluctuations dominate over the classical
instanton solution everywhere. Therefore, the Coleman instanton ($\chi =0$)
is never trustable$.$ The solution of equation $\left( \ref{35}\right) $ is 
\begin{equation}
\varrho _{uv}=\varrho _{b}\left( 1+\frac{1}{2}\frac{\nu }{\left( \left(
1+\beta \chi \right) \chi \right) ^{3/2}}+O\left( \frac{\nu ^{2}}{\chi ^{3}}%
\right) \right)  \label{77a}
\end{equation}%
and to the leading order the equations $\left( \ref{45}\right) $ and $\left( %
\ref{46}\right) $ simplify to%
\begin{equation}
\varphi_{uv}\left(\chi \right) \simeq -\varphi_{0}\,\left( \frac{1+\beta\,
\chi }{4\,\chi }-\frac{\nu ^{2}}{\left( 1+\beta\, \chi \right)^{2}\chi ^{2}}%
\right)  \label{78}
\end{equation}%
and 
\begin{equation}
V_{uv}\left( \chi \right) \simeq -\frac{4\,V_{bar}}{\beta}\left( \frac{1}{%
4\,\chi }-\frac{\nu^{2}}{\left( 1+\beta\, \chi \right)^{2}\,\chi ^{2}}\right) \,.
\label{79}
\end{equation}%
It follows from here that only if 
\begin{equation}
\chi >4\,\nu ^{2}  \label{80}
\end{equation}%
both $\varphi _{uv}$ and $V_{uv}$ are negative and hence the tunneling
becomes possible. This lower bound on $\chi $ is in complete agreement with
the bound $\left( \ref{54}\right) $ obtained by inspecting the relevance of
the infrared cutoff scale. Because of this bound the minimal value of the
potential in the center of the bubble must be always larger than 
\begin{equation}
V_{c}\simeq -\frac{32\,\varphi_{0}^{4}}{\sigma ^{2}}\,.  \label{81}
\end{equation}
Let us consider the $\chi_{\varepsilon }$ instanton for which 
\begin{equation}
V_{uv}\left( \chi _{\varepsilon }\right) =\varepsilon\, V_{c}\,,  \label{82}
\end{equation}%
where $\varepsilon <1.$ We obtain the corresponding $\chi _{\varepsilon }$
substituting expansion $\left( \ref{79}\right) $ into (\ref{82}), where we
keep only the first term,%
\begin{equation}
\chi _{\varepsilon }\simeq \frac{\sigma^{2}\,\lambda _{-}}{128\,\varepsilon }\,.
\label{82aa}
\end{equation}%
The size of the bubble is given by 
\begin{equation}
\varrho_{0}^{2}\,\varphi _{0}^{2}\simeq \frac{\sigma ^{2}}{16\,\varepsilon }%
\left( 1+\frac{\sigma ^{2}\,\lambda _{+}}{128\,\varepsilon }+O\left( \frac{%
\varepsilon}{\lambda_{-}}\right) \right) .  \label{82bb}
\end{equation}%
and this size is always larger than some minimal size, that is, 
\begin{equation}
\varrho _{0}\geq \frac{\sigma }{4\,\varphi _{0}},  \label{82cc}
\end{equation}%
irrespectively of the value of $\lambda _{-}>1.$ This resolves the problem
of small size instantons for the steep potentials. To calculate the action
we have to substitute $\left( \ref{82aa}\right) $ in $\left( \ref{57b}%
\right) .$ As a result one gets%
\begin{equation}
S\left( \chi_{\varepsilon }\right) \simeq \frac{\pi^{2}\sigma ^{2}}{%
16\,\varepsilon }\left[ 1+\frac{\sigma^{2}\,\lambda_{+}}{128\,\varepsilon }+%
\frac{1}{3}\left(\frac{\sigma^{2}\,\lambda_{+}}{128\,\varepsilon }\right)^{2}
+O\left( \frac{\varepsilon}{\lambda_{-}}\right) \right] \,.
\label{82dd}
\end{equation}%
Notice that both the bubble size or the action do not depend on $\lambda
_{-} $ in the leading order and therefore we can take a limit $\lambda
_{-}\rightarrow \infty $ (see Fig. \ref{Figure3}), when all $\lambda _{-}$%
-dependent corrections proportional to $\varepsilon /\lambda _{-}$ vanish.
Although all instantons contribute to the vacuum decay rate, that major
contribution to this rate for the unbounded potential and the potentials
with the second very deep minimum $V_{\min }<V_{c}$ (see curve (a) in Fig. %
\ref{Figure3}) give instantons with the minimal possible size and action,
corresponding to $\varepsilon \simeq 1,$ so that%
\begin{equation}
\Gamma \sim \varphi_{0}^{4}\,\exp \left[ -\frac{\pi ^{2}\sigma ^{2}}{16}%
\right] \,.  \label{83}
\end{equation}%
One has to stress that in this case we are working on the border of
semiclassical approximation and the emerging bubbles contain only one
quantum. Therefore equation $\left( \ref{83}\right) $ gives only a very rough
estimate for the probability of decay. To determine the leading contribution
to the false vacuum decay rate in case when the second minimum of depth $%
V_{\min }$ is above $V_{c}$ (see curve (b) in Fig. \ref{Figure3}) one has to
use the formulae $\left( \ref{82bb}\right) $ and $\left( \ref{82dd}\right) $
with $\varepsilon $ determined by equation $V_{\min }\simeq \varepsilon
V_{c}.$

\section{Conclusions}

In this paper we have analyzed the role of quantum fluctuations for the
classical $O\left( 4\right) $ solutions which describe the false vacuum
decay. It was shown that the quantum fluctuations induce the ultraviolet and
infrared cutoff scales, beyond which the classical instanton solutions
cannot be trusted anymore because the level of quantum fluctuations exceed
the contribution of the classical field. The cutoff scales are entirely
determined by the parameters characterizing the corresponding classical
solutions. This allows one to abandon the boundary condition imposed by
Coleman on the derivative of the field at $\rho=0$, which is outside
the range of validity of the approximation. The regularized solutions are
indeed not singular. As a result there emerges the whole spectrum of new
instantons, which all contribute to the false vacuum decay. The largest
contribution comes from the instantons with the minimal possible action. In
many cases these instantons (up to small corrections) lead to the same
result as instanton with the Coleman boundary conditions. However, in several
important cases the Coleman approach fails. In particular, we have shown
that for the very steep unbounded from below potential, shown in Fig. \ref%
{Figure3}, the instanton with the Coleman boundary conditions would lead to
nearly instantaneous vacuum instability via formation of infinitely small
bubbles. To the contrary, as the size of the new instantons always exceeds
some minimal scale determined by the parameters of the original potential,
the decay probability obtained by these new instantons remains finite even
for case where the unbounded potential is very steep. This removes the
difficulties that were associated with the application of 
small instantons�. We have also checked that the results in the
method we suggest agrees with the standard method when they overlap.

\vspace*{1.0cm}

\textbf{Acknowledgments}

\bigskip

The work of V. M. was supported under Germany's Excellence
Strategy---EXC-2111---Grant No. 39081486.

The work of E. R. is partially supported by the Israeli Science Foundation
Center of Excellence. E. R. would also like to thank NHETC at Rutgers Physics
Department, the IHES and the CCPP at NYU for hospitality.

A. S. would like express a gratitude to the Racah Institute of Physics of
the Hebrew University of Jerusalem and Ludwig Maxmillian University of
Munich for the hospitality during his visits. The work of A. S. was
partially supported by RFBR grant No. 20-02-00411.

\end{document}